# Frontoparietal Connectivity Neurofeedback Training for Promotion of Working Memory: An fNIRS Study in Healthy Male Participants

Meiyun Xia[1,3], Pengfei Xu[1,3], Yuanbin Yang[5], Wenyu Jiang[6], Zehua Wang[1,3], Xiaolei Gu[1,3], Mingxi Yang[1,3], Deyu Li[1,3,4], Shuyu Li[1,3], Guijun Dong[7], Ling Wang[2,*], and Daifa Wang[1,3,*]

[1]School of Biological Science and Medical Engineering, Beihang University, Beijing, 100083, P. R. China
[2]College of Computer Science, Sichuan Normal University, Chengdu, 610101, P. R. China
[3]Beijing Advanced Innovation Center for Biomedical Engineering, Beihang University, Beijing, 100083, P. R. China
[4]State Key Laboratory of Virtual Reality Technology and System, Beihang University, Beijing, 100083, P. R. China
[5]Department of Rehabilitation Medicine, Wangjing Hospital, China Academy of Chinese Medical Sciences, Beijing, 100102, P. R. China
[6]Department of Neurological Rehabilitation, Guangxi Jiangbin Hospital, Nanning, 530021, P. R. China
[7]School of Sports and Health, Shandong Sport University, Jinan, 250102, China

Corresponding author: Ling Wang (lingwang@sicnu.edu.cn) and Daifa Wang (daifa.wang@buaa.edu.cn).

This work was supported in part by the National Key Research and Development Plan under Grant 20YFC2004200 and Grant 2018YFC2001700; in part by the National Major Scientific Research Instrument Development Project of China under Grant 81927804; in part by the National Natural Science Foundation of China (NSFC) under Grant 31771071, Grant 81972160, Grant 81622025, and Grant 61101008; in part by the Chinese Ministry of Justice Key Research Project on the Detoxification Theory in Justice Administration under Grant 19ZD07; in part by the Guangxi Medical and Health Foundation for Development and Application of Appropriate Technology under Grant S2018092; and in part by the Guangxi Natural Science Foundation under Grant 2016GXNSFAA380110.

**ABSTRACT** Neurofeedback cognitive training is a promising tool used to promote cognitive functions effectively and efficiently. In this study, we investigated a novel functional near-infrared spectroscopy (fNIRS)-based frontoparietal functional connectivity (FC) neurofeedback training paradigm related to working memory, involving healthy adults. Compared with conventional cognitive training studies, we chose the frontoparietal network, a key brain region for cognitive function modulation, as neurofeedback, yielding a strong targeting effect. In the experiment, 10 participants (test group) received three cognitive training sessions of 15 min using fNIRS-based frontoparietal FC as neurofeedback, and another 10 participants served as the control group. Frontoparietal FC was significantly increased in the test group ($p = 0.03$), and the cognitive functions (memory and attention) were significantly promoted compared with the control group (accuracy of 3-back test: $p = 0.0005$, reaction time of 3-back test: $p = 0.0009$). After additional validations on long-term training effect and on different patient populations, the proposed method exhibited considerable potential to be developed as a fast, effective, and widespread training approach for cognitive function enhancement.

## I. INTRODUCTION

Cognitive training has become an important approach to promote cognitive functions in healthy people or patients with various neurodevelopmental and neurodegenerative diseases [1]-[6]. However, conventional cognitive training (e.g., aerobic exercise [7] and engagements in social activities and everyday intellectual activities [8]) is usually lengthy. Neurofeedback techniques provide a novel way of promoting the effectiveness and applicability of cognitive training. Based on the visualization of the neurophysiological status, neurofeedback techniques enable the participant to directly attempt to regulate his/her brain activity. Furthermore, the participant can upregulate or downregulate the neural activation of targeted brain regions and improve the outcome of cognitive training [2], [9]-[13].

Currently, the main neuroimaging modalities used in neurofeedback include single-modality, e.g., electroencephalography (EEG, for a review of EEG-neurofeedback, refer to [14], [15]), functional magnetic resonance imaging (fMRI, for a review of fMRI-neurofeedback, refer to [16]-[18]), and functional near-infrared spectroscopy (fNIRS, for a review of fNRIS-neurofeedback, refer to [10], [19]), and multimodality, e.g., EEG-fMRI neurofeedback [20]-[22], EEG–fNIRS neurofeedback [23], [24], and fMRI–fNIRS neurofeedback [25], [26]. EEG has an inherently high-temporal resolution and captures the summation of postsynaptic potentials of thousands and millions of pyramidal neurons. It can measure real-time brain activation information and has been used as the neurofeedback signal [27]. The spatial resolution of EEG is relatively low, and it is sensitive to motion. In turn, fMRI can noninvasively capture the blood oxygen level-dependent signal changes in deep-brain regions and provide near-real-time hemodynamic neurofeedback with high-spatial resolution over the entire brain. Owing to this advantage, fMRI has been increasingly utilized in neurofeedback studies. For example, patients with Parkinson's disease and attention deficit hyperactivity disorder (ADHD) successfully learned how to increase brain activities in their motor-related or attention-related cortices [28], [29]. However, fMRI measurement is associated with a stringent imaging environment that limits the design of neurofeedback training experiments.

FNIRS is a relatively new neuroimaging technique that has become a useful tool for brain activity monitoring. This modality has been increasingly employed in neuropsychiatric research, including patients with schizophrenia, affective and anxiety disorders, as well as eating disorders and ADHD [10], [30]. fNIRS measures changes in the optical properties of brain tissues in the near-infrared (NIR) range to estimate fluctuations in the concentration of oxyhemoglobin (HbO) and deoxyhemoglobin (HbR) associated with neural activities [12], [13], [19], [31]-[38]. Although limited by the penetration depth of NIR light inside biological tissues, fNIRS provides the tradeoff balance between moderate temporal and spatial resolutions when imaging the brain and can simultaneously locate specific cortical regions and measure the hemodynamic signals over the entire brain cortex. Moreover, fNIRS measurements tolerate more head motions than EEG and fMRI do. This makes it possible to use fNIRS in more naturalistic environments/situations (e.g., allowing neural activity to be recorded during overt speeches, movements, and direct interactions with other persons). Furthermore, it permits investigation of populations that are more likely to show head motion abnormalities (e.g., neurological or psychiatric patients or infants [32]) and of situations that do not allow fMRI measurements (e.g., participants with ferromagnetic implants or claustrophobia) [39]. Additionally, owing to its simplicity and limited cost [33], [34], fNIRS is fit for repetitive measurements and is, thus, a practical and convenient tool for neurofeedback applications in the practical clinical and rehabilitation environment. In recent years, important progress has been made in related research based on fNIRS, e.g., acupuncture, peripheral stimulation, and cognitive training to improve cognitive ability [40]-[42]. Previous studies have shown that fNIRS-based neurofeedback training can manipulate the activation of the lateral orbitofrontal cortex and prefrontal cortex in healthy participants [31], [43] and the motor cortex in healthy participants and patients after stroke [12], [13]. Compared with single-modality imaging methods, multimodality methods can combine the advantages of single modalities and yield brain activity views with an unprecedented spatiotemporal resolution, e.g., high-spatiotemporal resolution in neuroimaging can be achieved with EEG–fMRI setups. However, strong electromagnetic interference and motion artifacts are important topics for EEG-fMRI analysis. In addition, the availability of fMRI for multiple repeat scans, functional region-based task training and practical limitations brought by experimental environment, etc., make it relatively difficult for fMRI-based single-modality and multimodality feedback training in clinical practice. The most pressing problem of multimodality imaging is that it requires dedicated analysis methods. Till date, there has been no gold standard analysis framework for metabolic views on brain activity [44].

Connectivity neurofeedback has been developing in recent years. It is well known that the brain functions of human beings are coherently controlled by multiple brain regions called brain networks [45], [46], and many cognitive functions, as well as psychiatric [47]-[50] and neurodegenerative [50], [51] diseases, are closely related to brain networks. Most of the research in this field is based on fMRI. For example, memory is proved to be closely related to the frontoparietal and the default mode networks, whereas attention is related to the dorsal attention network [46], [52]-[54]. Fukuda *et al.* [55] and Yamashita *et al.* [9] revealed that fMRI-based connectivity neurofeedback training can induce the aimed directional changes in functional connectivity (FC) between the left primary motor cortex and the left lateral parietal cortex. Kim *et al.* also reported that the brain activity-plus-connectivity neurofeedback based on fMRI can help heavy smokers to effectively regulate their psychological functions [56].

In this study, we proposed a novel fNIRS-based frontoparietal FC neurofeedback training paradigm related to working memory (WM) and investigated whether the proposed paradigm can manipulate the frontoparietal FC and effectively promote cognitive functions by using fNIRS-based frontoparietal FC as neurofeedback. If the answer provided by this study is "yes", then, a low-cost, easy to use, potentially portable, and robust to motion connectivity neurofeedback training strategy can probably be developed because of the aforementioned advantages of fNIRS. In the proposed paradigm, a Sternberg WM task [57], [58] was executed at first, and fNIRS signals of frontal and parietal regions were acquired simultaneously. Subsequently, a feedback score computed from the strength of the frontoparietal FC measured by using fNIRS was displayed as feedback. During the neurofeedback tasks, the participants in the test group were instructed to make the feedback score as high as possible.

In the paradigm, WM was adopted owing to its critical involvement in the execution of cognitive tasks by the brain, and the frontoparietal FC was chosen as the modulation target. Previous studies have revealed that WM-related cognitive functions are closely related to the frontoparietal brain network [46], [52], [59]-[64]. For example, patients with schizophrenia, Alzheimer's disease, or other related diseases encounter cognitive declines in attention and WM that are found to be closely related to the frontoparietal FC [52], [59], [60].

In this study, the results of three cognitive training sessions of 15 min showed that the proposed fNIRS-based connectivity neurofeedback training paradigm can significantly upregulate the frontoparietal FC and further promote related cognitive functions. This study preliminarily verified the feasibility and effectiveness of improving cognitive functions of young healthy participants based on fNIRS-based brain connectivity neurofeedback training.

## II. MATERIALS AND METHODS

### A. PARTICIPANTS

This study was approved by the local Ethics Committee of Beihang University. Twenty right-handed healthy men (aged $23.5\pm1.3$ years; 22–26 years) were recruited from Beihang University for this study. All participants were naive to psychology experiments. Written informed consents were obtained from all participants before the experiment. A participation fee was afforded. All participants are native Chinese speakers whose second language is English, and all of them had passed the College English Test 4 (CET-4) in China. All participants had no history of brain injury, neurological disease, or other serious medical conditions, with normal or corrected normal vision. The participants were randomly divided into two groups (10 participants in each group), namely, the test group and the control group. The test group received feedback scores as neurofeedback during the three connectivity neurofeedback training sessions of 15 min (referred to as connectivity neurofeedback (CNF) training, explained in detail in Section II C). And the control group did not receive feedback score during CNF training. There was no significant difference in memory capability between the two groups of participants (see Table I for intergroup $p$-value of Baseline). Participants were instructed not to take analgesics, anti-inflammatory drugs, caffeine, or any stimulant drinks for at least 6 h before the experiment.

### B. fNIRS DATA ACQUISITION

The fNIRS signals were acquired by using the NirScan system (Huichuang, China). The sampling rate was 13 Hz, and the wavelengths used were 740 nm and 850 nm. As shown in Fig. 1, the probes (16 sources and 16 detectors) were placed on the scalp according to the international 10–20 system [65], [66]. In total, 42 fNIRS channels were used, covering the prefrontal (1–22 channels), parietal (23–40 channels), and temporal (41, 42 channels) cortices (see Fig. 1(a)).

### C. EXPERIMENTAL PROTOCOL

The implicit and delayed feedback method was adopted [67]-[71]. As shown in Fig. 2(a), the experiment consisted of six sessions, one baseline estimation session (referred to as Baseline), three CNF training sessions (including three sessions referred to as T1, T2, and T3), and two followup evaluation training sessions (referred to as Evaluation, including two sessions referred to as WEEK1 and WEEK3). All participants conducted the same tasks during Baseline and Evaluation, including the cognitive training without CNF, resting state measurement, and behavioral testing. The Baseline and each session of the CNF training were conducted on separate days within the first week, and there was a recess of one day between the Baseline and CNF training. That is, Baseline was conducted on the first day, and T1, T2, and T3 were conducted on the $3^{rd}$, $4^{th}$, and $5^{th}$ days, respectively. Next, WEEK1 and WEEK3 were conducted one and three weeks after T3, respectively. After each training session, the participant received a 3 min resting state measurement during which the participant was instructed to stay relaxed and watch the "+" sign appearing in the center of the screen. After the resting state measurement, the participant rested for 5 min. In addition, the participants need to undergo behavioral tests after the resting state of Baseline, T3, WEEK1, and WEEK3 (see Fig. 2(a)). During behavioral testing, the participant took part in a 3-back (~2 min) test, a psychomotor vigilance test (PVT, ~3 min), and a color-word stroop test (CWST, ~2 min) sequentially and rested for 5 min between two consecutive tests. Detailed procedures of the behavior testing are introduced in Section IIF. There were 25 trials (12–13 min) in each training session. During the CNF training, fNIRS-based CNF was provided to the participants in the test group at the end of each trial in the form of a feedback score appearing on the screen. The feedback score represented the strength of the participant's frontoparietal FC in this trial. Before the experiment, participants were instructed to try their best to increase the feedback scores during the experiment and were informed that their monetary reward was positively related to the feedback scores. The only difference between the experimental protocols of the two groups was whether the participants were provided with feedback scores during the CNF training.

The cognitive training paradigm was based on the verbal WM task (i.e., the Sternberg task, the frontal and parietal lobes are more involved in the execution). Each trial of a training session with neurofeedback consisted of six phases: a remind phase (3 s), a memory phase (2 s), a retention phase (10 s), an inquiry phase (2 s), a neurofeedback phase (2 s), and a rest phase (8 s). In the control group, the neurofeedback phase was missing, whereas the rest phase lasted for 10 s. In the remind phase, a 3 s countdown appeared in the center of the screen to remind the participants of the onset of a new trial. In the memory phase, six nonrepetitive English letters appeared on the screen, and the participant was instructed to memorize them within 2 s. The letters then disappeared, and the retention phase began. A "+" sign appeared in the center of the screen during which the participant needed to maintain the six nonrepetitive letters in memory for 10 s. The inquiry phase followed. A random letter appeared in the center of the screen, and the participant was instructed to a) judge whether this letter was among the six nonrepetitive letters in the memory phase and b) choose "yes/no" (by pressing "1/2" on the keyboard) according to his/her judgment. If the participant did not press any button on the keyboard in the inquiry phase, the judgment for this trial was regarded as erroneous. For participants in the test group, the feedback score appeared on the screen during the neurofeedback phase. A rest phase of 8 s followed before the onset of the subsequent trial. The probability that the six nonrepetitive letters displayed in the memory phase appeared in the inquiry phase was controlled to be 50%. To achieve this, 12 nonrepetitive letters were first randomly selected from the 26 letters in the English alphabet. Six nonrepetitive letters were then randomly selected from these 12 letters and displayed in the memory phase, and one letter was randomly selected from these 12 letters and displayed in the inquiry phase.

### D. RELATION BETWEEN HbO AND FRONTOPARIETAL FC Z-VALUE

The fNIRS data collected during the memory and retention phases (time window: 0–12 s with a delay of 2 s) were used in each trial to compute the frontoparietal FC and the feedback score. To reduce the physiological noise caused by heart beating, respiration, and other physiological processes, the recorded fNIRS signals were filtered with cut-off frequencies of 0.2 and 0.6 Hz [72]. We then extracted channels with significant responses. Next, the concentration changes of HbO and HbR were calculated according to the modified Beer–Lambert law. The frontoparietal FC values were then calculated from the HbO signals measured from the prefrontal and parietal cortices. The calculation steps were as follows.

First, region-averaged HbO signals corresponding to the prefrontal cortex ($x_A$, averaged HbO signal) and parietal cortex ($x_B$, averaged HbO signal) were calculated. Second, the Pearson's correlation coefficient ($r$) between $x_A$ and $x_B$ was calculated as follows:

$$r = \frac{\sum_{k=1}^{K}\left[x_A(k) - \overline{x_A}\right]\left[x_B(k) - \overline{x_B}\right]}{\sqrt{\sum_{k=1}^{K}\left[x_A(k) - \overline{x_A}\right]^2 \left[x_B(k) - \overline{x_B}\right]^2}}, \quad (1)$$

where $K$ is the number of values in $x_A$ and $x_B$, $x_A(k)$ and $x_B(k)$ are the $k^{th}$ values in $x_A$ and $x_B$, respectively, and $\overline{x_A}$ and $\overline{x_B}$ are the mean values of $x_A$ and $x_B$, respectively. Third, to observe the changes of the frontoparietal FC, a Fisher $r$-$z$ transform was performed on the calculated Pearson's correlation coefficient, and the $z$-value representing the frontoparietal FC value was calculated according to

$$z = \frac{1}{2}\ln\frac{1+r}{1-r}. \quad (2)$$

### E. RELATION BETWEEN Z VALUE AND FEEDBACK SCORE

The feedback score in each trial was calculated from the $z$-value measured during the memory and retention phases (time window: 0–12 s). Regarding the hemodynamic delay, we delayed the time window by 2 s, which is consistent with previous neurofeedback studies [43], [53], [65]. The feedback score of the $i^{th}$ trial was calculated as

$$Score_i = \frac{50(z_i + 3SD - z_{base})}{3SD} \quad (0 \leq Score_i \leq 100). \quad (3)$$

Here, $z_i$ is the frontoparietal FC $z$-value of the $i^{th}$ trial, and $z_{base}$ and $SD$ are respectively the mean and standard deviation of the $z$ values during Baseline. The feedback score provided the participant with the following information: the baseline performance corresponded to 50 scores, and a monetary reward was provided if the feedback score of the current trial exceeded 50 (i.e., the frontoparietal FC was higher than the baseline level). Scores that dropped below 0 or exceeded 100 were kept at 0 or 100, respectively. The online signal processing and all visual presentations in the experimental protocol were performed using MATLAB (R2016a, MathWorks, Natick, MA, USA), and the Psychtoolbox was used for visual presentations.

### F. BEHAVIORAL TESTING

The effects of the proposed neurofeedback training on cognitive functions were evaluated through behavioral tests. The classical 3-back test was designed to evaluate the changes in WM [73], [74]. 3-back is a classic memory paradigm that can be used to test the effect of training. In order to avoid the influence of proficiency caused by multiple training and verification, we chose the 3-back instead of the Sternberg task for the memory test. The accuracy and reaction time of the participants in the 3-back test were used as the primary outcome of the behavioral test.

Additionally, to investigate whether the proposed neurofeedback training affected other cognitive abilities beyond WM, the PVT and CWST were conducted. The PVT evaluates the ability to fix the attention, and CWST is a response inhibition test that evaluates the ability of the participant to inhibit inappropriate responses under certain conditions [9]. Similar to WM, the ability to fix the attention and the ability to inhibit inappropriate responses are also modulated by the frontoparietal brain network. The reaction time of the PVT and CWST was used to evaluate the possible migratory aptitude of the proposed neurofeedback training on other cognitive abilities. The experimental procedures in the behavioral testing were as follows.

#### 1) 3-BACK TEST

As shown in Fig. 2(d), during the 3-back test, a series of letters appeared on the screen in sequence. Each letter lasted 2 s, and the interval between the appearances of two sequential letters was 1 s (a "+" sign appeared on the screen during this interval). In the 3-back task, $n = 3$, and the participant began the assessment from the $4^{th}$ $((n+1)^{th})$ letter. Specifically, the participant assessed whether the $4^{th}$ letter was the same as the $1^{st}$ letter (4 - $n$ = 1, where $n$ = 3), $5^{th}$ letter was the same as the $2^{nd}$ letter (5 - $n$ = 2, where $n$ = 3), and so on. If two letters were the same, the participant pressed "1" on the keyboard; otherwise, "2" was pressed. Twenty-five trials were executed, and the mean accuracy and mean reaction time were calculated as the outcome of the 3-back test.

#### 2) PVT

As shown in Fig. 2(e), when a trial started, there was a white "+" sign at the center of the black screen. The participant was instructed to keep watching the screen in anticipation of the stimulus. When the stimulus appeared, the white "+" sign suddenly changed to red, and the participant needed to press button "1" promptly. If the participant successfully pressed button "1" within 3 s after the appearance of the stimulus, the red "+" sign changed to its original white color immediately after the action. Otherwise, the red "+" sign automatically changed to a white color after 3 s. Participants were instructed to attend 10 PVT trials, and the between-trial intervals ranged from 5 to 15 s. The time interval from the appearance of the stimulus to the moment when button "1" was pressed was the reaction time of each trial. The final PVT outcome was the averaged reaction time of the 10 trials.

3) CWST

As shown in Fig. 2(f), one of the three words, namely, "Red", "Yellow" and "Green" appeared on the screen randomly for 2 s. If the color of the word matched the meaning of the word (e.g., the color of the word "Red" was red), the participant pressed button "1"; otherwise "2" was pressed. Each participant took part in 30 trials, and the averaged reaction time (the time from the appearance of the word to the instant the button was pressed) over 30 trials was used as the result of the CWST.

## III. STATISTICAL ANALYSIS

One-way repeated measures ANOVA implemented in MATLAB was used to test the frontoparietal FC z-values of the test and control groups, following a false discovery rate post hoc pairwise t-test. Two sample t-tests were used to examine whether there were any differences between the two groups on cognitive abilities and behavioral performance. For data that did not meet normality, the Mann–Whitney U test was performed.

To further investigate the modulating effect of the CNF training on bilateral frontoparietal FCs, the changes of the left and right frontoparietal FCs' z-values were analyzed. Additionally, changes in the frontoparietal FC at the resting state were analyzed using the same statistical methods as those applied on the aforementioned task state data.

Additionally, to investigate whether the CNF training affected the FCs between the untargeted brain regions, HbO signals measured from the temporal cortex were analyzed. The changes of the intragroup (within the test group) and intergroup bilateral frontotemporal and temporal-parietal FCs were calculated and analyzed using the same statistical methods adopted in the analysis of the frontoparietal FC mentioned earlier. In all statistical analyses, a p-value of less than 0.05 was considered to be significant.

## IV. RESULTS

The main evaluation indicators of our training were the frontoparietal FC z-value and outcome of the behavioral tests (accuracy and reaction time). The group-wise (mean ± SD) main evaluation indicators in all experimental sessions are summarized in Table I. Furthermore, the HbO concentration changes in the prefrontal and parietal lobes of the two groups are shown in Fig. 3. Detailed data analysis results are provided in the following subsections.

### A. BEHAVIORAL PERFORMANCE

For the participants in the test group, the 3-back accuracy and reaction time had statistically significant differences from pre-training to post-training (3-back accuracy: $F_{3,27} = 35.83$ and $p < 0.001$; 3-back reaction time: $F_{3,27} = 24.32$ and $p < 0.001$). The 3-back accuracy (mean ± SD) significantly increased from 65 ± 11% to 81 ± 10% ($p = 0.00019$), and the 3-back reaction time significantly decreased from 1.35 ± 0.20 s to 1.14 ± 0.18 s ($p = 0.00006$) from Baseline to T3 (see Figs. 4(a) and 4(b)). Following WEEK1, the 3-back accuracy decreased to 75 ± 9% and the 3-back reaction time increased to 1.23 ± 0.19 s. Both outcomes were significantly different from the values measured in Baseline (accuracy: $p = 0.0004$; reaction time: $p = 0.0003$). After WEEK3, the 3-back accuracy were still significantly different from the baseline values ($p = 0.008$), whereas the 3-back reaction time became insignificant with respect to the baseline values ($p = 0.46$).

There were no significant differences between the test and control groups in either the 3-back accuracy ($p = 0.30$) or 3-back reaction time ($p = 0.99$) before the CNF training. Figs. 4(c) and 4(d) show the relative post-CNF training changes with respect to Baseline (relative changes = post-CNF training value - baseline value) for the 3-back accuracy and reaction time. After the CNF training, the 3-back accuracy increased significantly ($p = 0.0005$) and 3-back reaction time decreased significantly ($p = 0.0009$) in the test group compared with the control group. Such intergroup differences became insignificant after WEEK1 (3-back accuracy: $p = 0.12$; 3-back reaction time: $p = 0.39$) and WEEK3 (3-back accuracy: $p = 0.19$; 3-back reaction time: $p = 0.81$).

The PVT reaction times of the test group before and after training were statistically significant (test group: $F_{3,27} = 3.21$ and $p = 0.039$; control group: $F_{3,27} = 0.13$ and $p = 0.94$). Within the test group, the PVT reaction time decreased significantly ($p = 0.0054$) from 0.41 ± 0.06 s to 0.38 ± 0.05 s from Baseline to the post-CNF training level (see Fig. 4(e)). This decrease became insignificant after WEEK1 ($p = 0.21$) and WEEK3 ($p = 0.78$). There was no significant intergroup difference ($p = 0.67$) on the PVT reaction time in Baseline. After the CNF training, the relative changes of the PVT reaction time did not decrease significantly in the test group compared with the control group (T3: $p = 0.07$, WEEK1: $p = 0.16$, and WEEK3: $p = 0.88$, see Fig. 4(f)).

The CWST reaction times of the test group before and after training were not statistically significant (test group: $F_{3,27} = 0.37$ and $p = 0.78$, control group: $F_{3,27} = 0.53$ and $p = 0.67$). The CWST reaction time of the test group did not change significantly from Baseline to the post-CNF training level (see Fig. 4(g)). As shown in Fig. 4(h), there were no significant differences in the relative changes of the CWST reaction time after the CNF training between the test and control groups.

### B. TASK STATE FRONTOPARIETAL FC

Fig. 5 shows the changes in the frontoparietal FC z-value throughout the experiment (test group: $F_{5,45} = 3.15$ and $p = 0.016$; control group: $F_{5,45} = 0.83$ and $p = 0.53$). The z-values of the test group increased significantly from 0.91 ± 0.30 in Baseline to 1.21 ± 0.36 in T3 ($p = 0.03$). The significance of such an increase maintained in WEEK1 ($p = 0.03$), but disappeared in WEEK3 ($p = 0.86$). There were no significant intergroup differences in the frontoparietal FC z-value measured in Baseline ($p = 0.45$). In T3, the frontoparietal FC z-values in the test group were significantly higher than those in the control group ($p = 0.03$). These intergroup differences were insignificant in WEEK1 ($p = 0.08$) and WEEK3 ($p = 0.70$).

As shown in Fig. 6, we separately compared the changes in the bilateral frontoparietal FCs z-value (left frontoparietal FC of test group: $F_{5,45} = 3.02$ and $p = 0.020$; left frontoparietal

FC of control group: $F_{5,45} = 1.40$ and $p = 0.24$; right frontoparietal FC of test group: $F_{5,45} = 3.37$ and $p = 0.011$; right frontoparietal FC of control group: $F_{5,45} = 2.07$ and $p = 0.087$). It was found that within the test group, both the left and right frontoparietal FCs $z$-values increased from Baseline to T3, and the increase was more significant on the left side (left: $p = 0.014$; right: $p = 0.078$). In WEEK1, left frontoparietal FCs of the test group were still significantly different from the baseline values ($p = 0.04$).

In Baseline, there were no significant intergroup differences in bilateral frontoparietal FCs (left: $p = 0.64$; right: $p = 0.88$). In T3, bilateral frontoparietal FCs in the test group were significantly higher than those in the control group, and the significance was higher on the left side (left: $p = 0.016$; right: $p = 0.022$). In WEEK1 and WEEK3, there were no significant intergroup differences on either the left or right frontoparietal FC.

Fig. 7 shows the linear correlations between the z values of frontoparietal FC and 3 back accuracy. There is a positive correlation between frontoparietal FC and 3 back accuracy ($r = 0.83$), while there are negative relations between frontoparietal FC and 3 back reaction time ($r = -0.93$) and between frontoparietal FC and PVT reaction time ($r = -0.93$). The correlation coefficient of frontoparietal FC versus CWST reaction time is -0.11.

### C. RESTING STATE FRONTOPARIETAL FC

The fNIRS data measured from 30s to 150s during the resting state measurement were utilized for further analysis. The resting state bilateral frontoparietal FCs in the test group did not significantly increase from Baseline to T3 (left: $p = 0.83$; right: $p = 0.59$). Additionally, there were no significant differences in resting state bilateral frontoparietal FCs between the test and control groups, in either baseline or T3 (Baseline-left: $p = 0.6$; Baseline-right $p = 0.72$; T3-left: $p = 0.15$; T3-right: $p = 0.57$). Although there was no significant difference between the resting-state frontoparietal FCs before and after CNF training, it was worth noting that the resting-state $z$-values of the frontoparietal FC showed an upward trend during the three training sessions.

### D. TEMPORAL-LOBE-RELATED FCs

The regulation effect of fNIRS-based CNF training on the frontoparietal FC may radiate to other related FCs through the brain networks. Therefore, HbO signals were recorded from the temporal cortex in all experimental sessions to investigate the possible diversion effect of the CNF training regulation on the bilateral frontotemporal and temporal-parietal FCs. Results showed that within the test group, neither the frontotemporal nor the temporal-parietal FCs ($z$-value) changed significantly after the CNF training (in T3). Furthermore, there were no significant intergroup differences on either the frontotemporal or temporal-parietal FCs ($z$-value) after the CNF training.

### V. DISCUSSION

In this study, we proposed a novel fNIRS-based frontoparietal FC neurofeedback training paradigm related to WM. We investigated whether the proposed method can effectively regulate the frontoparietal FC and promote cognitive functions. Twenty healthy participants took part in the experiment. Ten participants received fNIRS-based CNF in the cognitive training, whereas the other 10 participants did not. Results showed that the frontoparietal FC was significantly upregulated after training, and the related cognitive performance was significantly promoted with short training time in the case of the participants who received fNIRS-based CNF. Moreover, there were significant post-training differences among participants who received/did not receive fNIRS-based CNF regarding the frontoparietal FC and cognitive performance. These results demonstrated that the proposed fNIRS-based CNF training is a promising approach for the upregulation of the frontoparietal brain network of healthy people and for the improvement of their cognitive performance. Interestingly, with the increase of CNF training times, the strength of frontoparietal FC of the test group showed a statistically linear upward trend (slope = 0.10), while after training, it showed a linear downward trend (slope = -0.16). We were interested in what was the maximum threshold of the frontoparietal FC strength and how long it would last as the training sessions increased and the followup time extended. Further research with longer training time and repeated measures of cognitive abilities after 6 or 12 months, or even longer is necessary.

The left side of the frontoparietal FC in the test group was significantly increased after the CNF training. This significant difference was maintained in WEEK1. Several studies have reported significant bilateral neural activation in both prefrontal and parietal cortices during the encoding, maintenance, and retrieval of the WM information [75], [76]. A recent study conducted by Baker *et al.* further revealed that compared with the visuospatial WM tasks that relied mainly on the neural activation of the right prefrontal cortex, the verbal WM tasks mainly activated the left prefrontal cortex [77]. In another early study, D'Esposito *et al.* analyzed the results from 20 WM-related fMRI/positron emission tomography studies and found that spatial WM tasks (similar to those in Baker's study [77]) exhibited greater activations in the right prefrontal cortex, whereas many nonspatial WM tasks (similar to those in this study) exhibited greater activation in the left prefrontal cortex [78]. Our results showed that the left frontoparietal FC of the test group was enhanced by using the proposed verbal WM task-related-paradigm, and the outcomes were consistent with previous findings. The results also showed that the neurofeedback training based on fNIRS frontoparietal FC can prolong the training effect of the left connection (to WEEK1).

Compared with the changes in task-state brain connectivity, the changes in the intrinsic resting state brain connectivity would more adequately reflect the effectiveness of cognitive training in the improvement of cognitive functions. In this study, we measured the resting state frontoparietal FC values for both hemispheres before and after the proposed neurofeedback training. Although there was no significant difference in the bilateral frontoparietal FCs before and after training, it was worth noting that the bilateral frontoparietal FCs showed an upward trend during the three training sessions. Certain previous studies have found that long-term cognitive training may change the resting state brain networks. For example, it was reported that after a period of WM-task-based cognitive training, the resting state frontoparietal FCs of the participants were altered [43], [79]. In the future, it is necessary for us to increase the training duration and number of sessions as well as the length of the followup period to validate the long-term training effects of the proposed paradigm.

Simple cognitive training has been argued for the lack of an effective diversion effect from targeted to untargeted cognitive regions [3]. To investigate whether the proposed fNIRS-based CNF training yielded a diversion effect, two other cognitive abilities regulated by the frontoparietal FC, i.e., the attention focusing ability and the inappropriate response inhabitation ability were tested before and after the CNF training. Results showed that the attention focusing ability (PVT reaction time) of the test group was significantly improved after the CNF training ($p = 0.0054$). This suggested that the proposed paradigm may yield a good diversion effect on the untargeted cognitive regions. Although the inappropriate response inhabitation ability (CWST reaction time) of the participants was not significantly improved after the training, we found that the reaction time of CWST was decreased after the CNF training, and the influence of the training paradigm time on the inappropriate response inhabitation ability should be further explored in the future.

The prefrontal and parietal cortices are known to be related to many other functional regions in the brain. As a result, we inferred that many related regions and their FCs may also be affected by the proposed neurofeedback training. Previous studies have reported that in addition to the prefrontal and parietal cortices, the temporal cortex plays an important role in the regulation of the process of WM [80], [81]. Therefore, in our study, the changes in the frontotemporal and temporal-parietal FCs were analyzed before and after the CNF training. Results showed that there were no significant changes in the frontotemporal FC within the test group after the CNF training. The reason for these results may also be attributed to the relatively short duration (three training sessions of 15 min) of the proposed CNF training, which was not long enough to induce significant changes in the indirectly regulated (untargeted) brain regions. Similar to our study, Fukuda *et al.* conducted an fMRI-based CNF study and found that their neurofeedback training (4 days) resulted in significant changes in the FC between targeted regions, but led to insignificant changes in the brain networks beyond the targeted regions [55]. Till date, it is still uncertain whether the CNF training affects the FCs beyond the targeted regions. We consider that it does and infer that the main reason why significant and positive changes were not observed in Fukuda's and our study is likely attributed to the relatively short duration of training.

To investigate the retainability of the proposed paradigm, two followup evaluations, WEEK1 and WEEK3, were conducted on all participants, one and three weeks after the CNF training, respectively. Results showed that with only three fNIRS-based frontoparietal CNF training sessions of 15 min, the training effects can be maintained for a week. In the future, if the proposed paradigm is applied as the routine training for cognitive enhancement, we are optimistic that with the increase of training frequency, the training effect can be prolonged for several months or even longer. This hypothesis will be studied in our future research.

Cognitive training that enhances frontoparietal FC makes the changes in the HbO concentration in the two brain regions more synergistic. Cognitive training is similar to physical training. Weight (HbO concentration changes in the frontal and parietal lobes) may not have a significant difference before and after training, but the body shape (FC) becomes more graceful and more coordinated.

The effectiveness of macro-control on global FC between two brain areas will provide a theoretical and experimental basis for more precise manipulation of the connections between localized brain subregions inside these areas. In this study, the prefrontal and parietal cortices were our regions-of-interest, and the global frontoparietal FC was provided as neurofeedback in the proposed cognitive training. The experimental results showed the importance of enhancing global frontoparietal FC for the improvement of cognitive functions. Further, our results on the bilateral global frontoparietal FCs during task and resting states validated the critical role of the left prefrontal cortex in the verbal WM tasks. Therefore, we believe that with more accurate brain region segmentation and data analysis methods, more precise localization of brain regions related to cognitive tasks can be obtained, thus providing feasible, refined, and more specific regulations among localized brain subregions.

Additionally, the effects of superficial signals must be considered. This study focused on the changes in the overall functional connections between the frontal and parietal lobes and did not consider the influence of superficial hemodynamic signals. In addition, a large number of existing neurofeedback studies based on fNIRS have achieved relevant results, but none of these studies has performed special processing on superficial signals. Given that fNIRS is highly sensitive to scalp hemodynamic fluctuations, in future neurofeedback studies, the effects of superficial signals on neurofeedback training effect are worth exploring further.

Several limitations associated with the present study should be noted. This study adopted three 15 min cognitive training sessions and followed up to the third week after the training. The training cycle was relatively short, and the number of training sessions was relatively small. Further research with longer training time and repeated measures of cognitive abilities after 6 or 12 months is necessary. The training cycle and followup time must be extended to further explore the impact of the training time on the proposed training paradigm, verify the sustainability of the training effect of participants, and explore the extent of the impact of this paradigm on other cognitive abilities. In addition, the study involved a control group with no feedback, and similar studies have previously been reported. In order to further improve the experimental design, no feedback group, true feedback group and sham feedback group will be set in future research work. Moreover, only male participants were recruited. Whether gender causes differences in training outcomes is worth exploring. Behavioral, biochemical, and physiological data in animals demonstrate that the gonadal steroid hormones, estrogen and progesterone, affect behavior and modulate neuronal activity [82]. These hormones can affect cognitive functions and affective state. A previous study also revealed that the activity in reward-related brain regions was both modulated by menstrual cycle phase and correlated with gonadal steroid hormone levels [82]. In the future, the effectiveness of the proposed paradigm on female participants will be investigated and the possible gender effect will be discussed. Previous studies have shown that the prefrontal and parietal lobes are more involved in the execution of the Sternberg task than the temporal lobes. Therefore, the frontoparietal FC was selected as the feedback target in this study. Of course, the temporal cortex also played an important role in the regulation of the process of WM. We will choose appropriate tasks to verify the feasibility of regulating temporal-parietal and frontotemporal connections in future studies.

## VI. CONCLUSIONS

In this study, we proposed a novel WM task-related, fNIRS-based frontoparietal connectivity neurofeedback training paradigm and verified its capability in manipulating the frontoparietal FC and improving the cognitive abilities within limited training times. The results showed that the proposed method can effectively upregulate the frontoparietal brain network and promote memory cognitive abilities only in three training sessions of 15 min. With further validations on different populations and brain networks, the proposed method shows potential to be developed as a fast, effective, and extensively used training tool for cognitive enhancement in the future.

## VII. DISCLOSURES

The authors declare no conflicts of interest related to this article.

## ACKNOWLEDGMENT

We would like to thank Editage (www.editage.cn) for English language editing.

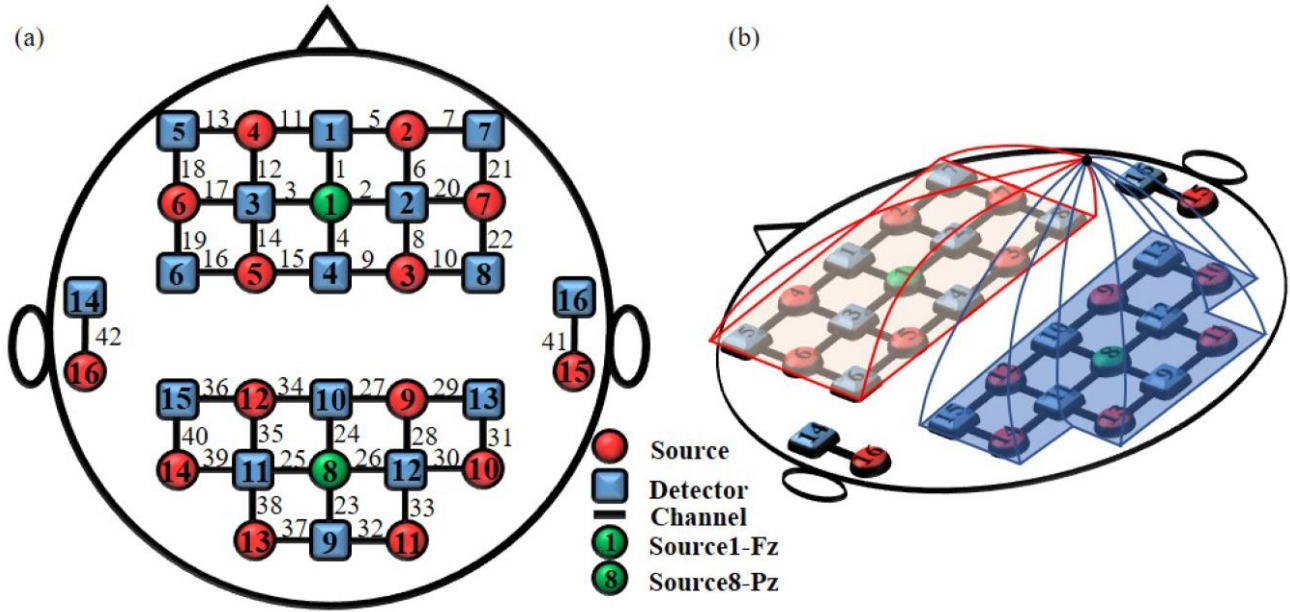

**FIGURE 1.** Neurofeedback setting. (a) A 42-channel functional near-infrared spectroscopy (fNIRS) probe set covering the prefrontal, parietal, and temporal regions. The red circles indicate fNIRS sources, blue rectangles indicate fNIRS detectors, and black lines indicate fNIRS channels. The sources (in green color) are placed on the Fz and Pz positions according to the international 10–20 system. (b) Frontoparietal functional connectivity (FC). It is the brain connection between the prefrontal and parietal cortical regions. That is, oxyhemoglobin (HbO) signals measured from different channels are region-averaged (red shaded area represents channels in the prefrontal region and blue shaded area represents channels in the parietal region), and frontoparietal FC is the functional connectivity between the prefrontal region-averaged HbO signal and the parietal region-averaged HbO signal.

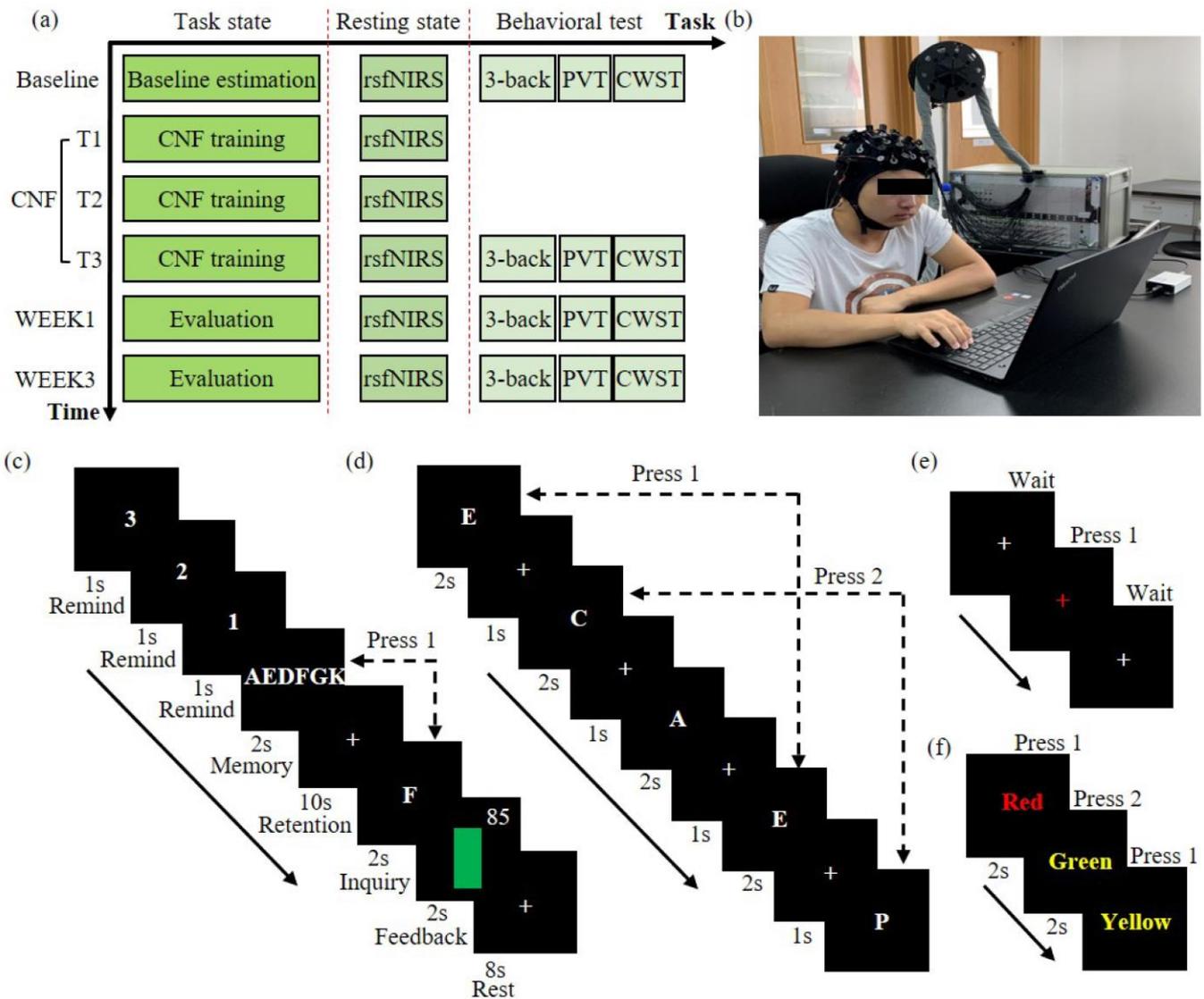

FIGURE 2. Experimental protocol. (a) Experimental workflow, including one baseline estimation session (Baseline) and three fNIRS-based connectivity neurofeedback training sessions (CNF training, including three sessions referred to as T1, T2, and T3) within the first week, followed by two followup evaluation sessions (Evaluation) after one week (WEEK1) and three weeks (WEEK3). Resting state fNIRS signals (rsfNIRS) were measured after each training session. Participants performed the behavioral tests after Baseline, T3, WEEK1, and WEEK3. (b) An actual image of the experimental setting. (c) A trial based on the Sternberg task with feedback score during the CNF training. (d), (e), and (f) 3-back test, psychomotor vigilance test (PVT) and color-word stroop test (CWST), respectively.

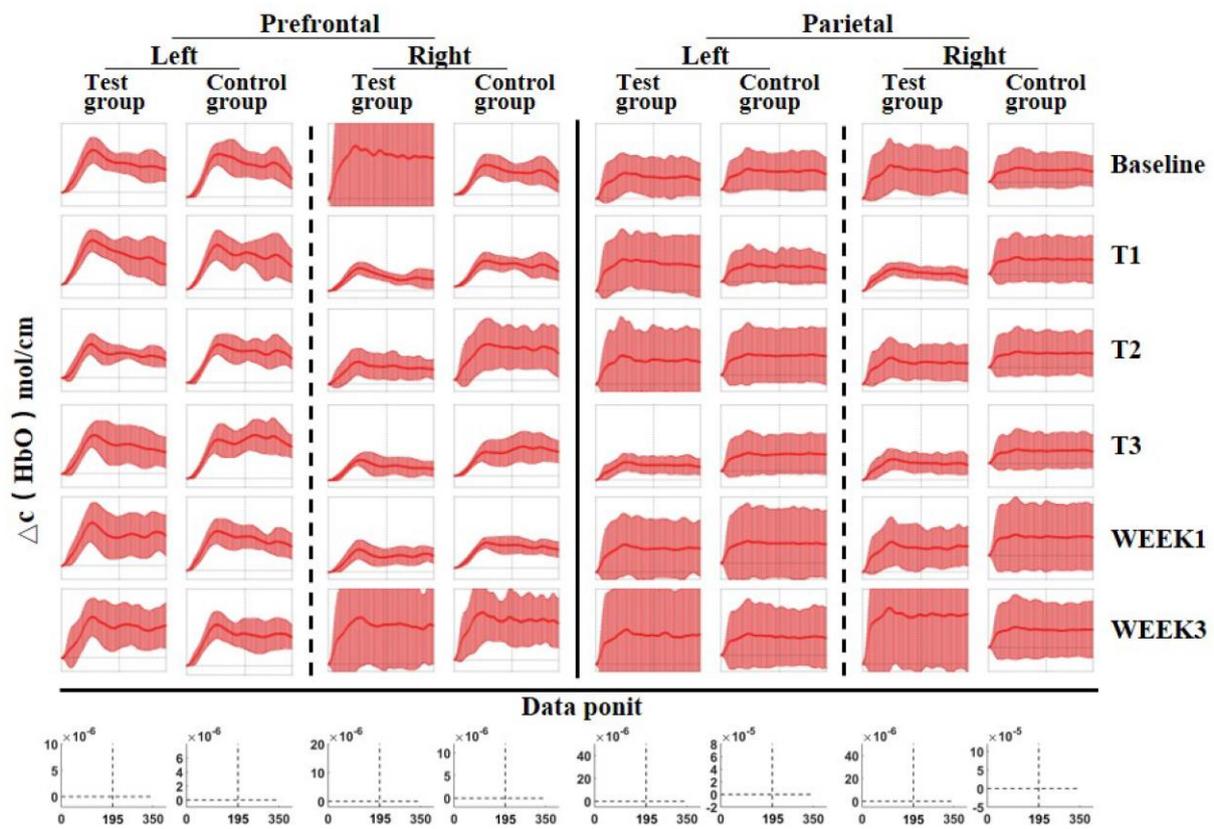

**FIGURE 3.** Means (thick red trace) and standard deviations of the HbO concentration changes in the prefrontal and parietal lobes of the test and control groups in the task state.

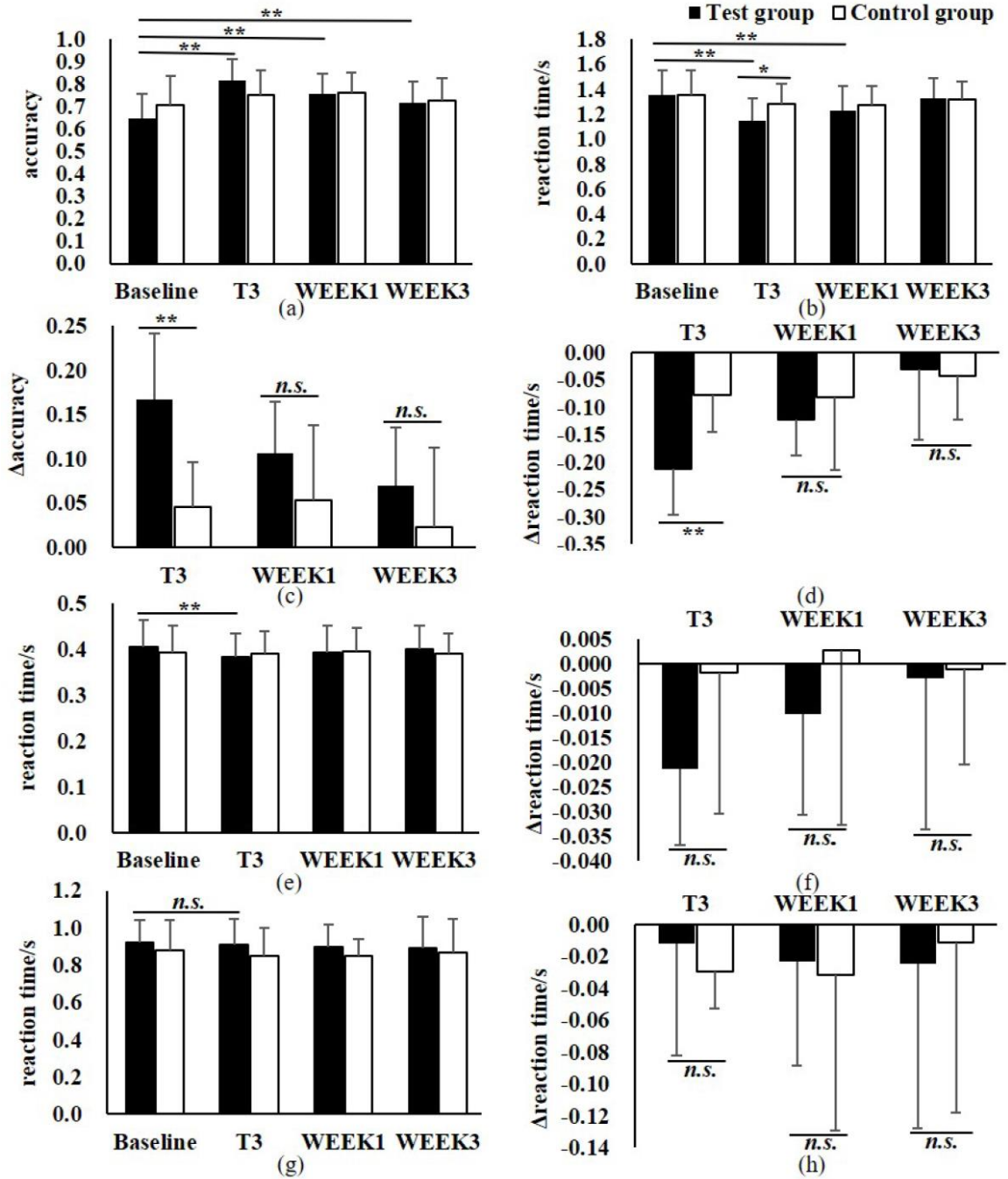

**FIGURE 4.** Behavioral testing results after the CNF training. Results of (a) 3-back accuracy and (b) 3-back reaction time. Relative changes in (c) 3-back accuracy, (d) 3-back reaction time. (e) PVT reaction time, (f) relative changes in PVT reaction time, (g) CWST reaction time and (h) relative changes in CWST reaction time (**: $p < 0.01$, *: $p < 0.05$, n.s.: not significant).

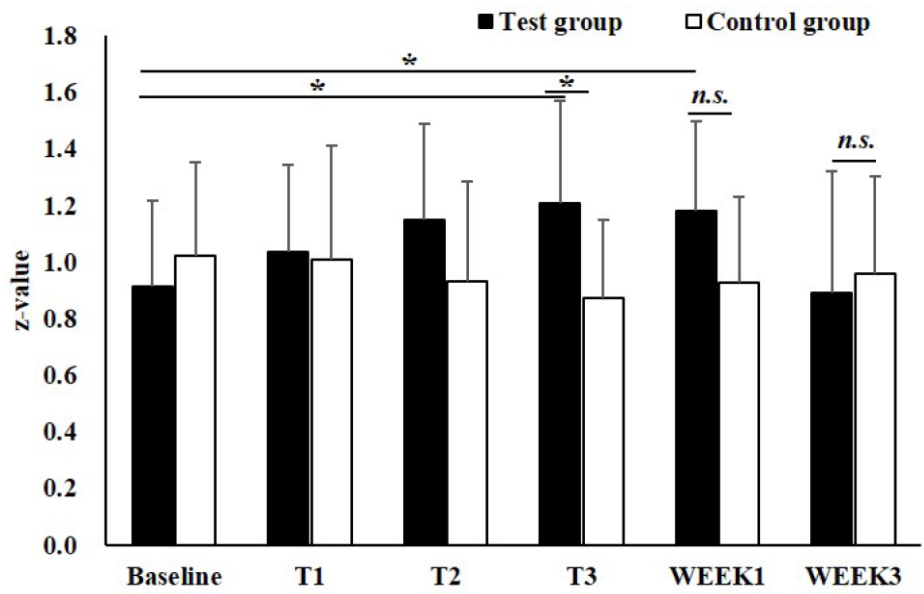

**FIGURE 5.** *z*-values of frontoparietal FC before and after CNF training (**: $p < 0.01$, *: $p < 0.05$, *n.s.*: not significant).

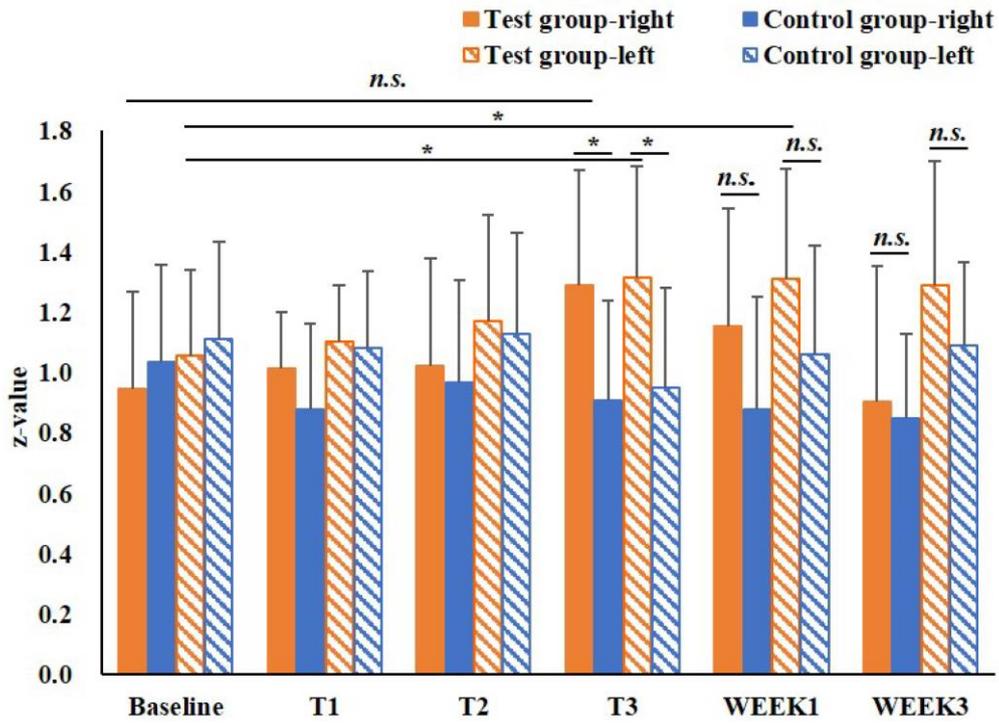

**FIGURE 6.** z-values of the task state bilateral frontoparietal FCs before and after the CNF training (\*\*: $p < 0.01$, \*: $p < 0.05$, n.s.: not significant).

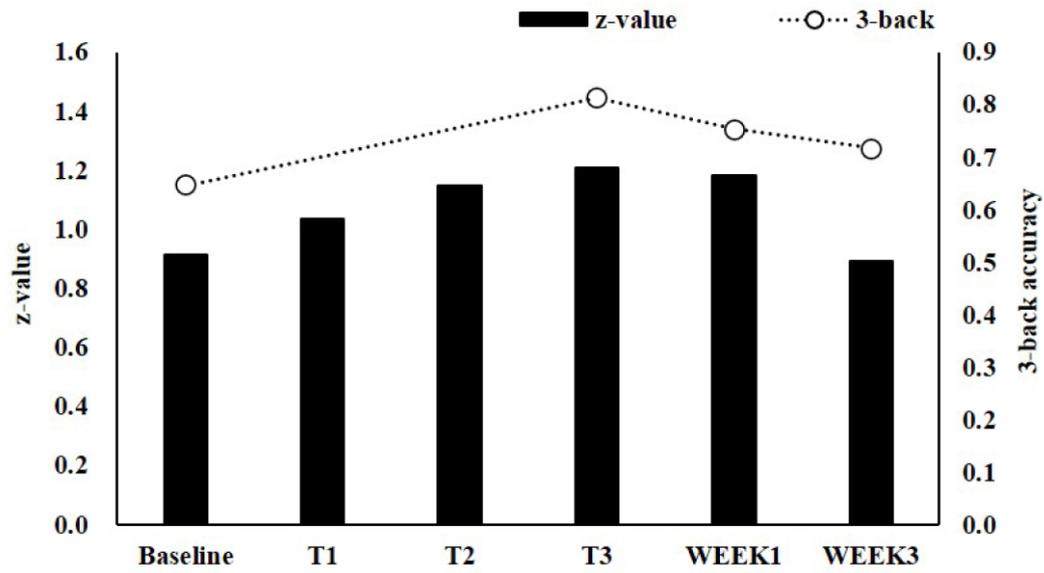

**FIGURE 7.** Changing trends in 3-back accuracy and the *z*-values of the frontoparietal functional connectivity within the test group before and after CNF training.

TABLE I
COGNITIVE TRAINING RESULTS

|  | Test group | | | | Control group | | | | Intergroup |
| --- | --- | --- | --- | --- | --- | --- | --- | --- | --- |
|  | Baseline | T3 | WEEK1 | WEEK3 | Baseline | T3 | WEEK1 | WEEK3 | P-value of Baseline |
| z-value | 0.91±0.30 | 1.21±0.36 | 1.18±0.31 | 0.89±0.43 | 1.02±0.33 | 0.87±0.28 | 0.93±0.30 | 0.96±0.34 | 0.45 |
| left z-value | 1.05±0.27 | 1.31±0.29 | 1.31±0.32 | 1.29±0.42 | 1.11±0.29 | 0.95±0.32 | 1.06±0.23 | 1.09±0.33 | 0.64 |
| right z-value | 1.01±0.32 | 1.31±0.38 | 1.15±0.39 | 0.90±0.45 | 1.03±0.32 | 0.91±0.33 | 0.88±0.37 | 0.85±0.28 | 0.88 |
| 3-back accuracy | 0.65±0.11 | 0.81±0.10 | 0.75±0.09 | 0.72±0.09 | 0.70±0.13 | 0.75±0.11 | 0.76±0.09 | 0.73±0.10 | 0.30 |
| 3-back reaction time | 1.35±0.20 | 1.14±0.18 | 1.23±0.19 | 1.32±0.16 | 1.36±0.19 | 1.28±0.17 | 1.27±0.15 | 1.31±0.15 | 0.99 |
| PVT reaction time | 0.40±0.06 | 0.38±0.05 | 0.39±0.06 | 0.40±0.05 | 0.39±0.06 | 0.39±0.05 | 0.40±0.05 | 0.39±0.04 | 0.67 |
| CWST reaction time | 0.92±0.13 | 0.91±0.14 | 0.89±0.12 | 0.89±0.16 | 0.88±0.16 | 0.85±0.15 | 0.85±0.09 | 0.87±0.18 | 0.56 |

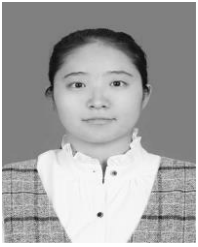

**MEIYUN XIA** received the B.S. degree in biomedical engineering from Yanshan University in 2016. She is currently pursuing the Ph.D. degree at Beihang University. Her research interests include fNIRS-based neurofeedback.

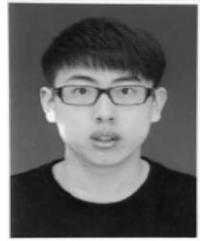

**MINGXI YANG** received the B.S degrees in biomedical engineering from China Medical University in 2018. He is currently pursuing the M.S. degree at Beihang University. His research interests include brain function evaluation by fNIRS.

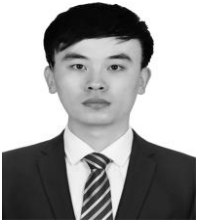

**PENGFEI XU** received the B.S. and M.S. degrees in biomedical engineering from Beihang University in 2017 and 2020, respectively. His research interests include fNIRS-based neurofeedback.

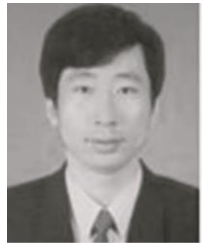

**DEYU LI** received the B.S. degree in engineering mechanics from the Chengdu University of Science and Technology in 1989, the M.S. degree in fluid mechanics from the University of Science and Technology of China in 1992, and the Ph.D. degree in biomedical engineering from Sichuan University in 2002. He has been involved with biomedical imaging and biomedical signal processing research for many years. He is currently a Professor with the School of Biological Science and Medical Engineering, Beihang University. His main research interests include biomedical imaging and image processing, biomedical signal processing and instrumentation, biomechanics, and rehabilitation engineering. He is a member of the Chinese Biomedical Engineering Society, a Council Member of the Biomedical Electronics Branch, Chinese Institute of Electronics, and a Council Member of the Beijing Biomedical Engineering Society

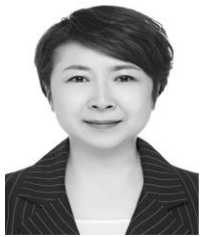

**Yuanbin Yang** received the B.S. degrees from Chongqing Medical University in 1988 and received the M.D. degrees from Capital Medical University in 2014. She is currently as a chief in Department of Rehabilitation Medicine in Wangjing Hospital of China Academy of Chinese Medical Sciences and an Associate Professor with the China Academy of Chinese Medical Sciences. Her research interests focus on rehabilitaion of neurological disease and tramatic brain injury.

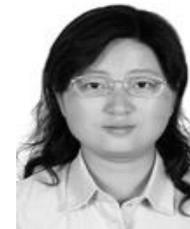

**Shuyu Li** received the B.S. degree in biomedical engineering from Capital Medical University, Beijing, China, in 1998, and the Ph.D. degree in Biophysics from Institute of Biophysics, Chinese Academy of Sciences, China, in 2003. From 2003 to 2005, she was a postdoctoral fellow at the National Laboratory of Pattern Recognition, Institute of Automation, Chinese Academy of Sciences, China. She is currently a Professor with the School of Biological Science and Medical Engineering at the Beihang University, Beijing, China. Her research interests span three major areas, including neuroimage analysis, progression of Alzheimer's disease, and the adolescent brain structural and functional development.

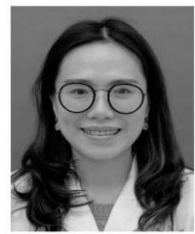

**Wenyu Jiang** received the M.M. and Ph.D. degrees in Nuerology at Guangxi medical University. She is currently an chief physician and the head of the Department of Neurological Rehabilitation, Guangxi Jiangbin Hospital. Her research interests include cognitive psychology, brain imaging, and rehabilitation.

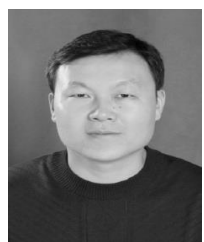

**GUIJUN DONG** received the PhD degree from the Chinese Academy of Sciences in 2006. He is currently a Professor and Vice Dean of the Institute of Sport and Health, Shandong Sport University. He is also the Vice President and Secretary General of the Sports Science Branch of Chinese Biophysical Society, Chairman of the Shandong Sports and Health Committee, Executive Director of the Shandong Gerontology and Geriatrics Society, Director of the Shandong Sports Society, Chief Expert of Sports detoxification in Shandong Province, and a member of drug detoxification Expert Group of Ministry of Justice of China. His main research area is the brain remodeling of exercise intervention substance addicts and the molecular mechanism of exercise intervention in osteoarthritis

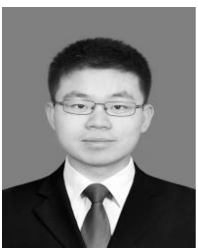

**ZEHUA WANG** received the B.S. and M.S. degrees in biomedical engineering from Beihang University in 2011 and 2014, respectively. He is currently pursuing the Ph.D. degree at Beihang University. His research interests include fNIRS-based neurofeedback.

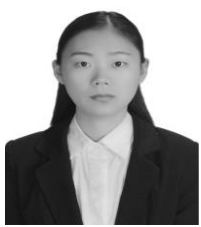

**XIAOLEI GU** received the B.S. degree in biomedical engineering from Changchun University of Science and Technology in 2018. She is currently pursuing the M.S. degree at Beihang University. Her research interests include neurofeedback based on fNIRS.

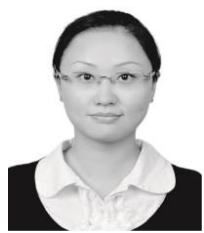

**LING WANG** received the B.S. degree in biomedical engineering from Sichuan University in 2005 and the Ph.D. degree in electronic engineering from the Chinese University of Hong Kong in 2010. She is currently an Assistant Professor with the College of Computer Science, Sichuan Normal University. Her research interests include fNIRS, brain cognitive science, and educational neuroscience.

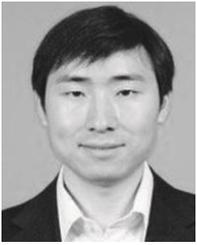

**DAIFA WANG** received the B.S. and Ph.D. degrees in biomedical engineering from Tsinghua University in 2005 and 2010, respectively. He is currently an Associate Professor with the School of Biological Science and Medical Engineering, Beihang University. His research interests include fNIRS, brain imaging, neuromodulation, and brain function evaluation.